\documentclass[]{emulateapj}

\usepackage{graphicx}
\usepackage[colorlinks=true,linkcolor=blue,citecolor=blue]{hyperref}

\shorttitle{Radio emission from a GRB cocoon}
\shortauthors{De Colle et al.}

\begin{document}


\title{Radio emission from the cocoon of a GRB jet: implications for relativistic supernovae and off-axis GRB emission
}


\author{Fabio De Colle\altaffilmark{1}, Pawan Kumar\altaffilmark{2}, David R. Aguilera-Dena\altaffilmark{3}}
\altaffiltext{1}{Instituto de Ciencias Nucleares, Universidad Nacional Aut{\'o}noma de M{\'e}xico, A. P. 70-543 04510 D. F. Mexico  \email{fabio@nucleares.unam.mx}}
\altaffiltext{2}{Department of Astronomy, University of Texas at Austin, Austin, TX 78712, USA}
\altaffiltext{3}{Argelander Institute for Astronomy, University of Bonn, Auf dem H\"ugel 71, 53121 Bonn, Germany}



\begin{abstract}
Relativistic supernovae constitute a sub-class of type Ic supernovae (SNe). Their non-thermal, radio emission differs notably from that of regular type Ic supernovae as they have a fast expansion speed (with velocities $\sim$ 0.6-0.8 c) which can not be explained by a ``standard'', spherical SN explosion but advocates for a quickly evolving, mildly relativistic ejecta associated with the SN. In this paper, we compute the synchrotron radiation emitted by the cocoon of a long gamma-ray burst jet (GRB). We show that the energy and velocity of the expanding cocoon, and the radio non-thermal light curves and spectra are consistent with those observed in relativistic SNe. Thus, the radio emission from this events is not coming from the SN shock front, but from the mildly relativistic cocoon produced by the passage of a GRB jet through the progenitor star. We also show that the cocoon radio emission dominates the GRB emission at early times for GRBs seen off-axis, and the flux can be larger at late times compared with on-axis GRBs if the cocoon energy is at least comparable with respect to the GRB energy.

\end{abstract}


\maketitle


\section{Introduction}

Gamma-ray bursts (GRBs) are pulses of high energy radiation emitted from collimated, highly relativistic jets 
\citep[e.g.,][]{kumar15}.  Long GRBs (LGRBs) are associated to type Ic supernovae (SNe), i.e. SNe produced by 
the collapse of massive, Wolf-Rayet progenitors \citep[e.g.,][]{cano17}. These are stars stripped of their 
hydrogen and helium envelope by strong winds \citep{yoon05} during the 
evolutionary phases preceding the core collapse. LGRBs are created during the stellar collapse and the
formation of a neutron star/black hole system 
(the so-called ``central engine'', CE hereafter; \citealt{usov92, woosley93}). 

While LGRBs are associated to a small fraction of type Ic SNe explosions ($\sim$ 1\%), 
also observations of low-luminosity GRBs, broad-lined SNe, relativistic SNe \citep{soderberg10, margutti14, chakraborti15} and super-luminous SNe  \citep[e.g.,][]{levan13, greiner15, inserra16,kann16,margutti17,coppejans18} indicate that matter is accelerated to large velocities, which is not easily explained by a ``standard'', neutrino-driven spherical supernova, requiring instead a CE, possibly similar to LGRBs \citep[e.g.,][]{lazzati12, nicholl17, sobacchi17, suzuki17a, suzuki17b}.

\begin{figure}
\centering
  \includegraphics[width=0.5\textwidth]{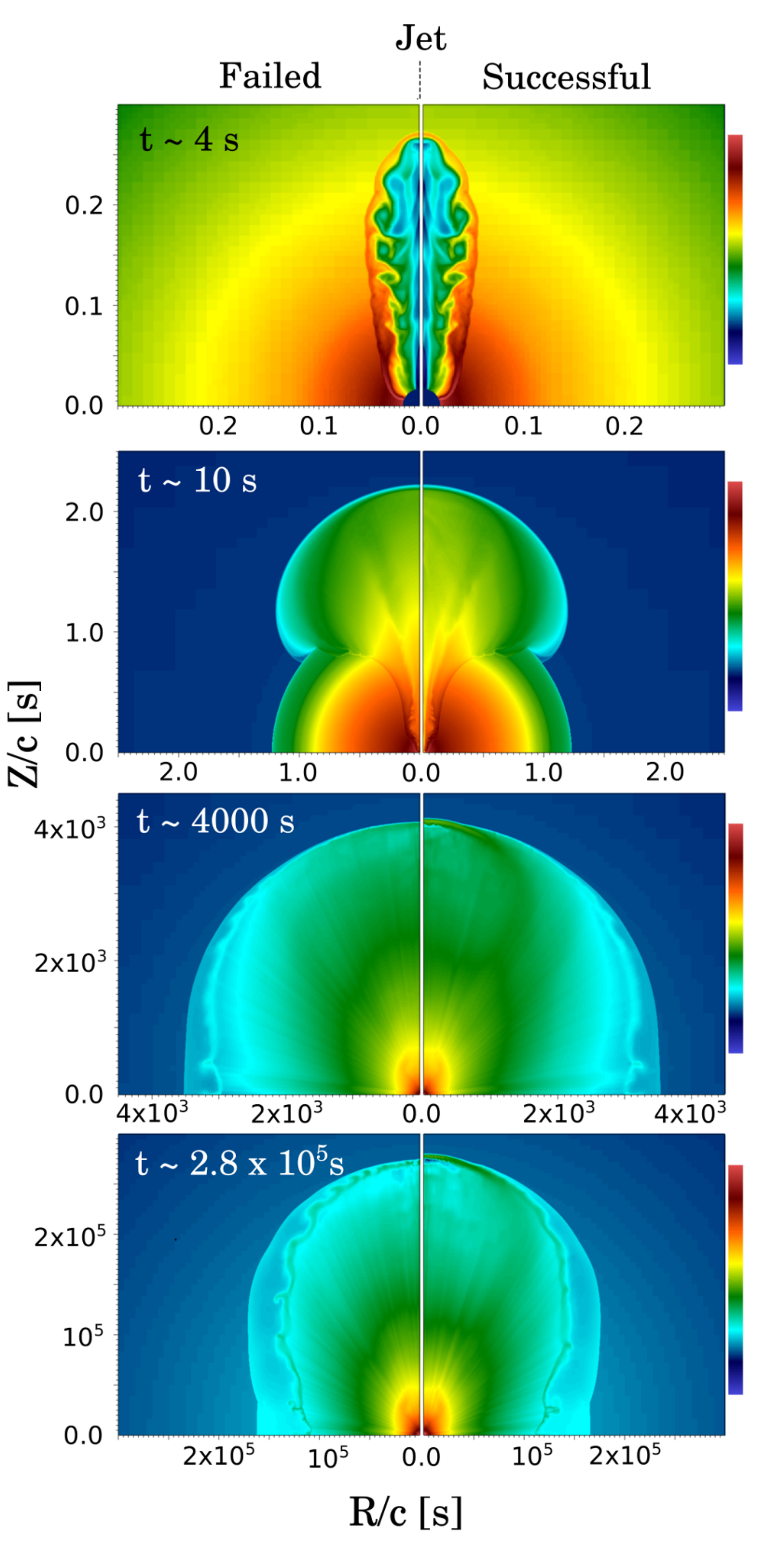}
\caption{Number density maps showing the dynamical evolution of a long gamma-ray burst jet. The color bar corresponds to the following density range (\emph{top} to \emph{bottom} panels, with larger densities in red and low densities in blue): (10$^{22}$, 10$^{30}$) cm$^{-3}$; (10$^{11}$, 10$^{30}$) cm$^{-3}$; (10$^3$, 10$^{22}$) cm$^{-3}$; (1,10$^{15}$) cm$^{-3}$. The jet is launched from an inner boundary located at a $r= 2 \times 10^8$~cm. It first moves with non-relativistic speed through the star (top panel), breaks out of the star and expands through the wind of the progenitor Wolf-Rayet star (central panels), forming an extended, nearly spherical cocoon (bottom panel) which expands at mildly relativistic speeds into the environment. Left panels show a ``failed'' jet, while right panels show a successful one. Differences between successful and ``failed'' jets can be seen only at late times in the region close to the jet axis (see, e.g. the \emph{bottom} panel), where the successful jet moves relativistically while the expanding shock wave produced by the ``failed'' jet begins to decelerate.}
\label{fig1} 
\end{figure}

The rare events that produce relativistic expansion velocities could represent only a small fraction of the cases in which a CE (and the associated jet) plays a role. In fact, 
CE-driven explosions have also been suggested as an alternative mechanism to neutrino energy deposition, to produce regular SNe \citep[see, e.g.,][and references therein]{bear17, piran17, soker17}.  Thus, understanding the observable signatures of central-engine driven SNe and their 
connection to GRBs may represent an important step in understanding the more general problem of SNe explosions itself.

Dozens of of SNe type Ibc have been observed in radio \citep[see, e.g., the review by][]{chevalier16}. 
In most cases SNe type Ibc present a synchrotron radio spectrum consistent with self-absorption and 
optically thin synchrotron emission at low and high frequencies respectively. 
The synchrotron radiation is emitted by electrons accelerated by the SN shock front, which therefore 
tracks the fastest and more energetic material. By analyzing the evolution of their spectrum in radio, 
it is possible to infer the physical parameters regulating the propagation of the shock wave associated with the expanding SN through the circumstellar material. 

Consistent with analytical models \citep[e.g.,][]{chevalier98}, observations show that the SN shock wave moves with nearly constant speed once it breaks out of the progenitor star, as $R\propto t^m$, where $m\lesssim 1$ and $R$ the position of the SN shock front. Observations show that the post-shock material moves with typical velocities $\sim 0.1$ c and energies $\sim$ $10^{46}$-$10^{48}$~erg.

Among type Ic SNe observed in radio frequencies, SN 2009bb and SN 2012ap show a peculiar behavior\footnote{The iPTF17cw broad-line type Ic SN is also a candidate for the class of relativistic SN \citep{corsi17}, but more observations are needed to confirm it.}. 

Their spectral evolution is in fact consistent with that of a decelerating, mildly relativistic shock with $R\propto t^m$ (with $m=0.8-0.9$) and $v_{\rm sh}\approx 0.6-0.8$~c, and an associated energy of  $\approx 10^{49}$~erg  (\citealt{soderberg10, chakraborti11, margutti14, chakraborti15, milisavljevic15}; see however \citealt{nakauchi15} who estimated a lower energy and velocity for SN 2009bb).

When the SN shock front breaks out of the star, it accelerates due to the large density gradients present in the stellar envelope \citep{sakurai60}. As a result, 
the expanding material will be strongly stratified, with most of the mass moving at (relatively) low velocities, and with a small amount of mass (close to the shock front) moving at relativistic speed. In terms of kinetic energy, \citet{tan01} showed that 
$E(>\Gamma\beta)\propto (\Gamma \beta)^{-5.1}$ in the newtonian limit (for shock waves moving through an $n=3$ politrope), and $E(>\Gamma\beta)\propto (\Gamma \beta)^{-1.1}$ in the ultra-relativistic limit. Thus, the determination of the energy producing the radio emission (which tracks the fast moving material), allows us to estimate the total energy of the spherical explosion. 
The energy of  $\approx 10^{49}$~erg associated to velocites $v_{\rm sh}\approx 0.6-0.8$~c in relativistic SNe, requires a spherical explosion with total energy $\gtrsim 10^{53}$~erg, which is several times larger than the most energetic SNe. Thus, as mentioned above, it has been suggested that these events are perhaps associated with a CE able to drive large energies at relativistic velocities.

In this paper we present numerical simulations of the expansion of a GRB jet through the progenitor star. 
We show that the non-thermal synchrotron emission produced by the cocoon shock front reproduces the radio emission observed in relativistic SNe. 
Furthermore, we show that the cocoon non-thermal radio emission can be 
of the same magnitude and in some cases larger than the emission from a GRB jet that is not pointing in our direction. Thus, the cocoon emission could be an important component to be considered for future observations of off-axis GRBs.

This paper is organized as follows: we present in Section 2 numerical simulations of the propagation of a GRB jet through a progenitor star, and the propagation of the associated cocoon. In section 3 we compute the radio emission and discuss the results. In Section 4 we present our conclusions.


\section{Cocoon dynamics}

We study the dynamical evolution of a GRB cocoon by running two-dimensional, axisymmetric simulations of the GRB jet as it expands through the progenitor star and its surrounding medium.
The simulations employ the adaptive mesh refinement, special relativistic, hydrodynamics code \emph{Mezcal} \citep{decolle12a}. We present here a brief description of the numerical simulations and refer the interested readers to \citet{decolle17}, where the numerical details of similar simulations were described 
more in detail.

As a progenitor star, we employ a 25 M$_\odot$ Wolf-Rayet pre-supernova stellar model (the E25 model of \citealt{heger00}).
The ambient medium density is given by $\rho=\dot{M}_w/4\pi r^2 v_w$, with $v_w=10^8$~cm s$^{-1}$ and $\dot{M}_w=2\times 10^{-6}$~M$_\odot$~yr$^{-1}$. The value of $\dot{M}_w$ was chosen to match the value inferred from observations of SN 2009bb \citep{soderberg10}.

In the simulations, a jet is launched from a spherical boundary located at $R=2\times 10^8$ cm. The jet is injected into the computational box from $t=0$~s to $t= t_{\rm jet}$, with a luminosity $L_{\rm jet}=2\times 10^{50}$~erg s$^{-1}$, a Lorentz factor $\Gamma_{\rm jet}=20$ and an opening angle $\theta_{\rm jet}=0.1$. 
We run two models. In the first one, $t_{\rm jet}=10$~s (model $t_{10}$ or ``successful'' jet model), while in the second $t_{\rm jet}=4$~s (model $t_{4}$ or ``failed'' jet model). The jets have the same luminosity thus differ in their total energy.

The computational box extends from 0 to 10$^{16}$~cm both along the $r-$ and $z-$ axis respectively\footnote{With respect to the simulations presented in \citet{decolle17}, the size of the box has been extended by a factor of $\sim 30$.}.
The AMR grid employs 60 $\times$ 60 cells with 26 levels of refinement, which corresponds to a resolution of $5\times 10^7$~cm. The extremely large number of levels of refinement employed in these simulations allow us to run them over $\sim 9$ orders of magnitude in space, and $\sim$ 26 orders of magnitude in density.

Figure \ref{fig1} shows the dynamical evolution of the jet and the cocoon. Once the jet is ejected from the central engine, it digs a hole through the star (see Figure \ref{fig1}, top panel). Due to the presence of the dense stellar material, the jet moves at non-relativistic velocities ($v\sim0.1$~c), depositing its kinetic energy into a cocoon, which provides extra collimation to the jet. In model $t_{10}$, the jet breaks out of the star, accelerates to its terminal velocity, and moves with nearly constant speed into the environment, while in model $t_4$ the jet is chocked inside the star. However, as most of the jet energy is deposited in the outer layers of the star, it breaks out of the star acquiring mildly relativistic velocities.

The main difference resulting from the evolution of the jets in the two models is the presence of a collimated component along the direction of propagation of the jet in model $t_{10}$, which is absent in the model $t_5$ (see Figure \ref{fig1}). On the other hand, the dynamical evolution and structure of the cocoons at large polar angles remain nearly identical for the full extent of the simulation.

The total energy in the cocoon is given by
\begin{equation}
 E_{\rm c} \lesssim L_{\rm jet} t_{\rm bo} = 5 \times 10^{50} L_{\rm jet, 50}\; t_{\rm bo, 5 \; \rm s}\;,
\end{equation}
where $t_{\rm bo}$ is the jet break out time (taken here equal to 5 s).
There are two cocoon components. The cocoon formed by stellar material shocked well before the jet broke out of the star expands slowly into the star (in the direction perpendicular to the direction of propagation of the jet), 
eventually breaking out of it on a timescale of $\sim$ 40 s. This ``inner'' cocoon contains most of the total cocoon energy.

On the other hand, the material which crossed the jet shock front a time 
$t\lesssim R_\perp/c_s \sim R\theta_j/c_s$ ($c_s$ being the sound speed of the shocked stellar material) before the jet breaks out, accelerates quickly engulfing the star in $\sim 15$~s, and then expands nearly spherically. 
This quickly expanding cocoon component contains a fraction of the total cocoon energy given by
\begin{equation}
 E_{\rm c} \approx L_{\rm jet} R\; \theta_j/c_s = 10^{49} L_{50} R_{11}\theta_{j,0.1} c_{s,9}^{-1} {\rm \;erg}\;.
\end{equation}

\begin{figure}
\centering
  \includegraphics[width=0.5\textwidth]{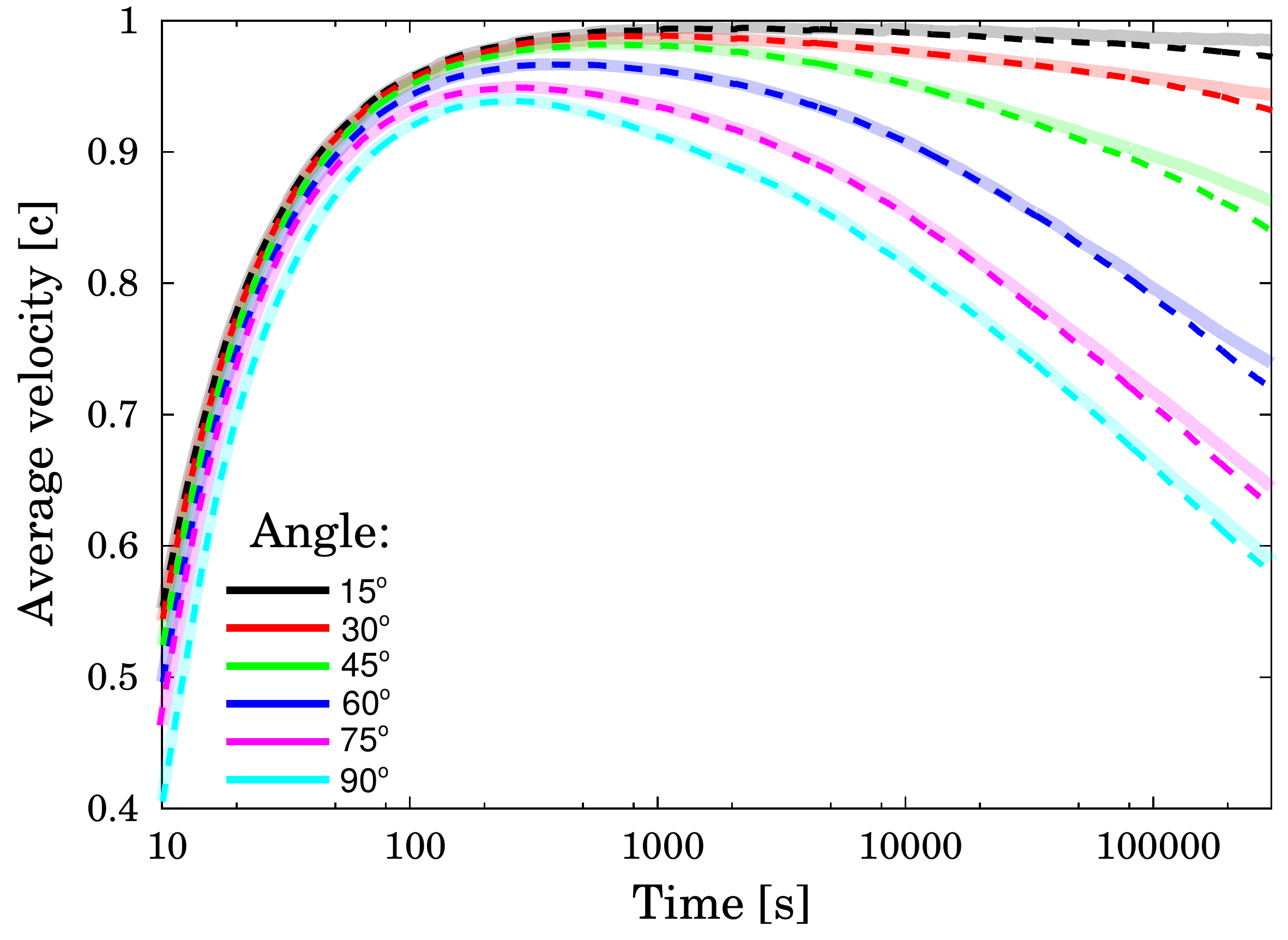}
\caption{Average cocoon shock velocity ($=\int_0^R \beta dt^\prime / t$) as a function of time (computed in the lab frame) for the successful (solid lines) and ``failed'' (dashed lines) jets respectively. The velocity is obtained by deriving the shock position $R_{\rm sh}(\theta,t)$ (computed from the numerical simulations) with respect of time, at different polar angles ($\theta=15^\circ, 30^\circ,45^\circ,60^\circ,75^\circ,90^\circ$, top to bottom lines).
}
\label{fig2} 
\end{figure}

Figure \ref{fig2} shows the time evolution of the cocoon shock velocity at different polar angles. The cocoon first accelerates to highly relativistic speeds at $t\sim 100-1000$~s. Then, while the material moving close to the jet channel continues to move at nearly constant speed, at large angles the cocoon decelerates to mildly relativistic velocities over timescales $t\sim 10^5$~s.
The evolution of the cocoon velocities of the failed and the successful jets are remarkably similar, although the cocoon produced by a failed GRB decelerates at a slightly faster rate.

At distances $\sim 10^{16}$~cm, the cocoon shock velocity becomes highly stratified along the polar ($\theta$) direction, ranging between 0.5~c and 0.9~c. The cocoon is also highly stratified along the radial direction, with a density profile decreasing as $\approx r^{-2}$, except along the jet channel which is instead filled by a low density, highly relativistic material in the successful jet model. Fits to the curves presented in Figure \ref{fig2} give $R\propto t^{-m}$, with $m=0.85-0.99$ for $\theta=90^\circ-15^\circ$ respectively.
 
\begin{figure}
\centering
  \includegraphics[width=0.5\textwidth]{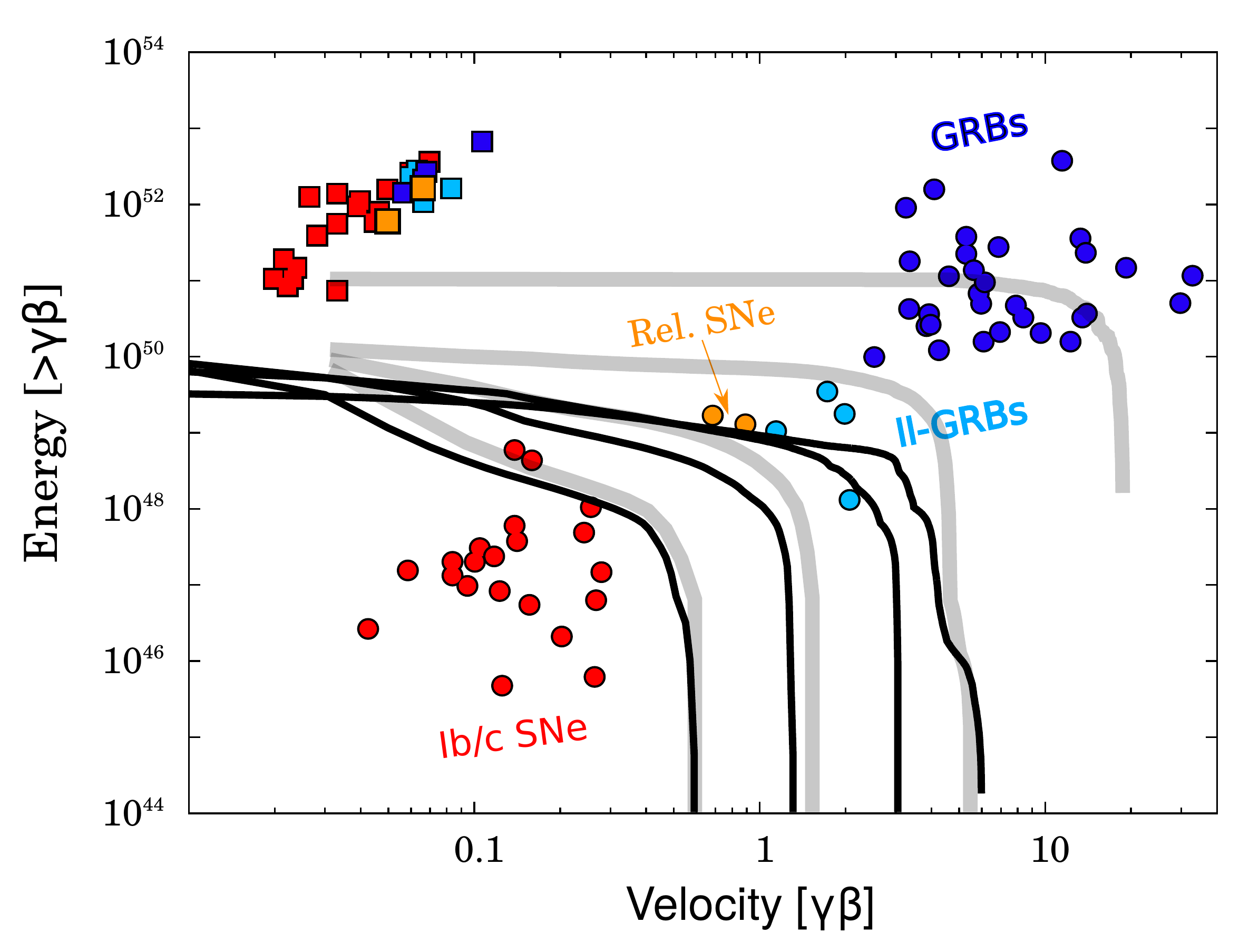}
\caption{Kinetic energy as a function of the ejecta velocity $\Gamma \beta$ for type Ic SNe observed in optical (red squares) and in radio (red circles), for ``standard'' GRBs (blue circles) and their associated SNe (blue squares) and for low-luminosity GRBs and relativistic SNe and their associated SNe (light blue circles and squares, respectively). The curves show the cocoon energy computed from the simulation of a successful (light gray) and failed jet (black) at different polar angles ($0^\circ$, $30^\circ$, $60^\circ$, $90^\circ$, from the fastest to the slowest curves respectively). Both models reproduce the energy of the relativistic SNe deduced from radio emission. Optical observations (squares) require a spherical SN component which is not included in our simulations. The figure is adapted from \citet{margutti14}.}
\label{fig3} 
\end{figure}

The cocoon energy distribution also strongly depends on the polar angle, and differs from the energy distribution observed in expanding spherical SNe and relativistic jets. 
Figure \ref{fig3} shows the kinetic energy (integrated over velocities $v > \Gamma\beta$) inferred from observations of type Ic SNe, GRBs and relativistic SNe. While typical type Ic-SNe can be explained by spherical symmetric explosions, injection of energy at large velocities must be considered to explain the energies observed in the GRBs and relativistic SNe. Figure \ref{fig3} clearly shows that the energy of the cocoon is consistent with the energy inferred from observations of relativistic SNe and low-luminosity GRBs. The successful GRB model has a larger energy at small polar angles (corresponding to energy launched towards larger polar angles from the collimated relativistic outflow). At angles $\gtrsim 50^\circ$ the energy distribution of the cocoon is nearly identical in the two cases considered.

\section{Synchrotron radiation from the cocoon}

\subsection{Methods}

We compute the synchrotron radiation emitted from the cocoon shock front by post-processing the results of the numerical simulations. As the simulation extends ``only'' to $10^{16}$~cm, we fit the velocity as a function of time and polar angle. Then, we use these extrapolated values for the velocity at later times when necessary in the calculation of the radiation.

Given the shock velocity as a function of time and polar angle (see, e.g., Figure \ref{fig2}) and the ambient density, we compute the post-shock density $\rho_{\rm ps}$, thermal energy density $e_{\rm ps}$ and velocity $v_{\rm ps}$. 
 We then assume that there is a non-thermal population of electrons (accelerated by the shock) with a distribution $N_e\propto \gamma_e^{-p}$, and with an energy density $e_{\rm acc}=\epsilon_e e_{\rm ps}$, i.e. given by a fraction $\epsilon_e$ of the post-shock thermal energy density. We also assume that the magnetic energy density is a fraction $\epsilon_B$ of the thermal energy, i.e. $B = \sqrt{8\pi \epsilon_B e_{\rm ps} }$.
 
To determine the observed synchrotron flux, we integrate (at fixed values of $t$, $\theta$) the radiation transfer equation through the post-shock region\footnote{We assume that the emission comes from the shocked wind, and neglect the emission due to the reverse shock.}. 
Assuming that the emitting region is uniform, the radiation transfer equation has the following solution
\begin{equation}
  I_{\nu^\prime} =  \frac{j_{\nu^\prime}}{\alpha_{\nu^\prime}} \left(1-e^{-\tau_{\nu^\prime}}\right)\;,
\label{eq:inu}
\end{equation}
where the proper frequency, $\nu^\prime$, is related to the observed frequency by
\begin{equation}
   \nu^\prime = \nu_{\rm obs} \gamma \left(1-\beta \cos\theta\right)\;.
\end{equation}

The flux is then computed from the specific intensity by integrating the equation
\begin{equation}
    F_\nu = \frac{1}{D^2} \int I_\nu dA \;.
\end{equation}

In equation \ref{eq:inu}, $\tau_{\nu^\prime}$ is the optical depth, while $j_{\nu^\prime}$ and $\alpha_{\nu^\prime}$ are the  specific emissivity and absorptivity respectively, defined as (\citealt{granot99}, see also \citealt{decolle12a, vaneerten12})\footnote{The cooling frequency is much larger than the radio frequencies for typical values of shocks velocity and density, so it is not considered here.}
\begin{eqnarray}
   j_{\nu^\prime} = 
0.88 \frac{64 q_e^3}{27 \pi m_e c^2} 
 \frac{(p-1)}{3 p-1} 
\frac{\xi_e n_{\rm ps} B_{\rm ps} }{\gamma^2 (1-\beta_\parallel)^2} \times \nonumber\\
\times \left\{ \begin{array}{cc} 
    \left(\nu^\prime/\nu_m\right)^{1/3} & \nu^\prime \leq \nu_m \\
    \left(\nu^\prime/\nu_m\right)^{(1-p)/2}  & \nu^\prime > \nu_m
  \end{array} \right.
\end{eqnarray}

\begin{eqnarray}
  \alpha_{\nu^\prime} = 
       \frac{\sqrt{3} q_e^3(p-1)(p+2)}{16 \pi m_e^2 c^2}  \frac{\xi_e n_{\rm ps} B_{\rm ps} \gamma(1-\beta_\parallel)}{\gamma_m \nu^{\prime,2}}\times  \nonumber\\
\times \left\{ \begin{array}{cc} 
    \left(\nu^\prime/\nu_m\right)^{1/3} & \nu^\prime \leq \nu_m \\
    \left(\nu^\prime/\nu_m\right)^{-p/2}  & \nu^\prime > \nu_m
  \end{array} \right.
\end{eqnarray}
with 
\begin{eqnarray}
   \nu_m = \frac{3 q_e \gamma_m^2 B}{4 \pi m_e c}\;, \qquad \beta_\parallel = \beta \cos\theta \;.
\end{eqnarray}

In these equations, $\beta_\parallel$ is the velocity parallel to the direction of the observer, $\xi_e$ is the fraction of post-shock electrons accelerated by the Fermi process (we assume $\xi=1$ in the calculations presented in this paper), $\gamma$ is the Lorentz factor of accelerated electrons, $\nu_m$ is the characteristic frequency corresponding to the minimum Lorentz factor of the accelerated electrons, and the other constants have their usual meaning.

Finally, we notice that at distances $r \gtrsim 10^{15}$ and for the mass-loss considered here ($\sim 10^{-6}$ M$_\odot$ yr$^{-1}$), free-free and Thomson scattering are negligible \citep[see, e.g.,][]{chevalier98}.

\subsection{Comparison with observations of SN 2009bb}

\begin{figure}
\centering
  \includegraphics[width=0.5\textwidth]{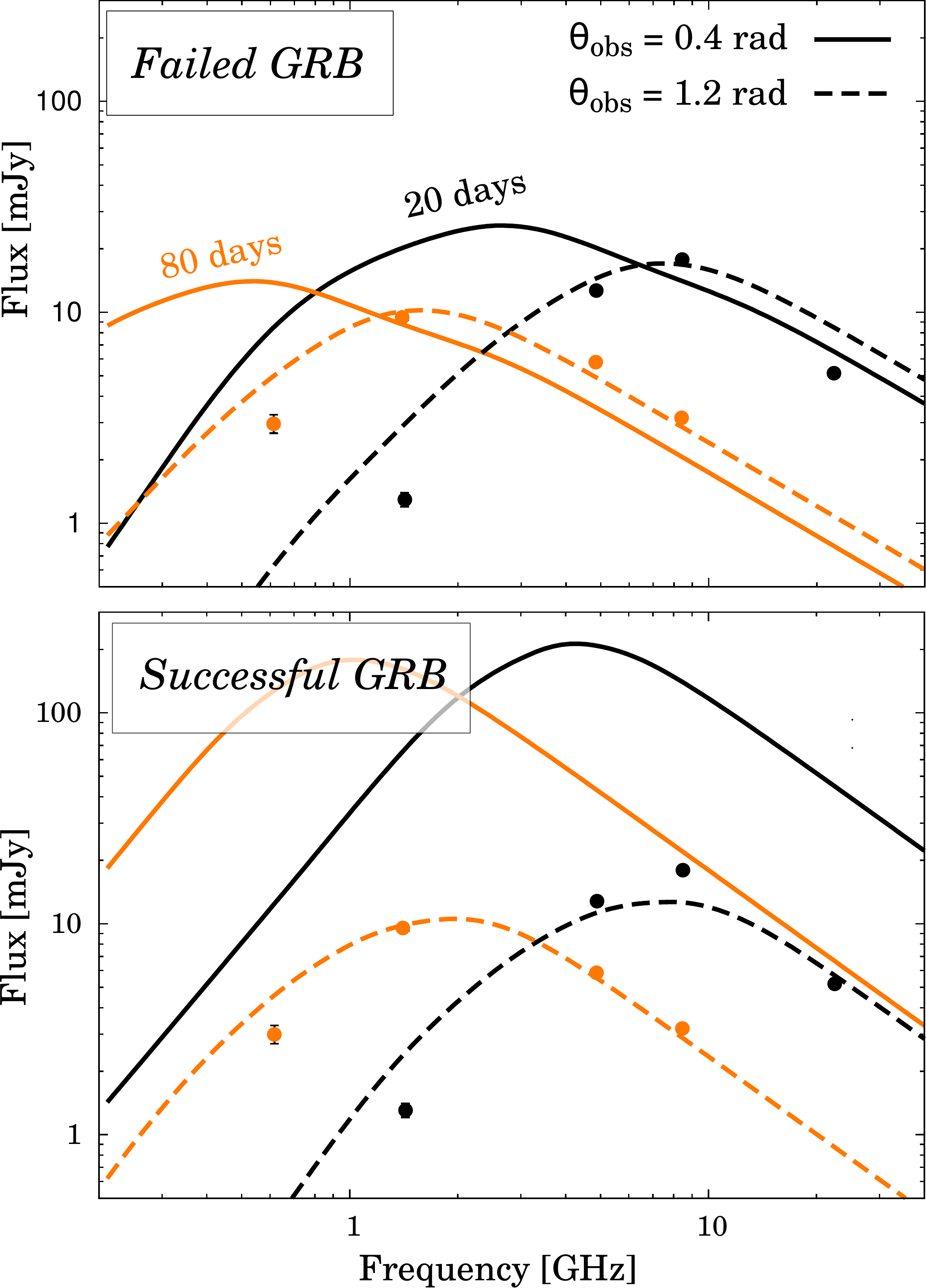}
\caption{Cocoon spectra 20 days (black full lines) and 100 days (red, dashed lines) after the SN explosion, compared with data from SN 2009bb. The curves correspond to different observer angles ($\theta_{\rm obs}=0.4, 0.8, \pi/2$ rad from the brightest to the dimmest curve).
The flux is computed at a distance of 40 Mpc. Observations of SN 2009bb \citep{soderberg10} at the same times are shown in the figure for comparison. Both models reproduce qualitatively the observations. 
}
\label{fig4} 
\end{figure}

The relativistic supernova SN 2009bb exploded on March 2009 and is located at $\sim 40$~Mpc in the nearby spiral galaxy NGC 3278. SN 2009bb has been classified as a broad lined type Ic SN, with photosphere velocities $\geq$~20000 km s$^{-1}$ and with a kinetic energy of $1.8\times10^{52}$~erg \citep{pignata11}.
This SN has been extensively observed at radio wavelengths with the VLA \citep{soderberg10}, VLBI \citep{bietenholz10} and GMRT \citep{ray14} spanning $\Delta t\sim$ 20-1000 days, and is about $\sim 100$ times more luminous than the ``average'' SN type Ibc. 
The spectrum is consistent with synchrotron emission, with the low-frequency part suppressed by synchrotron self-absorption.  The emission is well modeled by a shock with energy $\sim 10^{49}$~erg and a velocity $\sim 0.85\pm0.02$~c \citep{soderberg10}. 

As mentioned in the previous section, the energy and velocity of the expanding cocoon are of the same order of magnitude as those of relativistic supernovae.
Thus, it is expected that the cocoon non-thermal emission should be similar to the one observed in SN 2009bb. 

The synchrotron radiation emitted by the cocoon is presented in Figure \ref{fig4} at 20 and 80 days after the explosion. The best fit to the data at $\theta_{\rm obs} = 1.2$~rad is obtained for $\epsilon_B=0.27$, $\epsilon_e=0.019$ and $p=3.4$ in the successful jet model, and $\epsilon_B=0.25$, $\epsilon_e=0.05$ and $p=3$ in the failed jet model. In both cases, $\dot{M}_{\rm w} = 2\times 10^{-6}$~M$_\odot$ yr$^{-1}$.
The synthetic spectrum is also computed at $\theta_{\rm obs} = 0.4$~rad with the same parameters.

The synchrotron spectra show an optically thick (with $F_\nu \propto \nu^{5/2}$) and optically thin (with $F_\nu \propto \nu^{-(p-1)/2}$) component. As expected, the self-absorption frequency moves toward lower frequencies with time. 
Due to the decrease of the shock velocity with time, the peak flux also drops slightly 
with time. 

While in relativistic flows (e.g., GRBs) usually $\nu_m \gg \nu_a$ ($\nu_m$ is the characteristic frequency emitted by electrons accelerate with the minimum Lorentz factor $\gamma_m$), the opposite is true in non-relativistic flows (as $\nu_m\propto \epsilon_B^{1/2} \epsilon_e^2 e^{5/2} \chi_e^2 n^{-2}\propto \Gamma_{\rm sh}^3$ in relativistic and $\propto \beta_{\rm sh}^5$ in non-relativistic shocks respectively), in which case, $\nu_m \lesssim 10^9$~GHz. 

The cocoon is strongly asymmetric along the polar direction (see Figures \ref{fig1}, \ref{fig2}). As a consequence, the cocoon radio emission depends on the observing angle, increasing by $\sim$ one order of magnitude for observers located at $\sim 0.4$~rad with respect to observers located at $\theta_{\rm obs}=\pi/2$.
The emission from the cocoon is similar in the two models considered in this paper at large observing angles, differs when observed closer to the jet axis and is qualitatively consistent with the observations.
However, an off-axis jet could dominate the emission (at all observing angles) at larger times 
when its velocity has dropped to sub-relativistic speed and if its energy exceeds the energy in the cocoon (see the discussion in Section \ref{secgrb}).

\begin{figure}
\centering
  \includegraphics[width=0.48\textwidth]{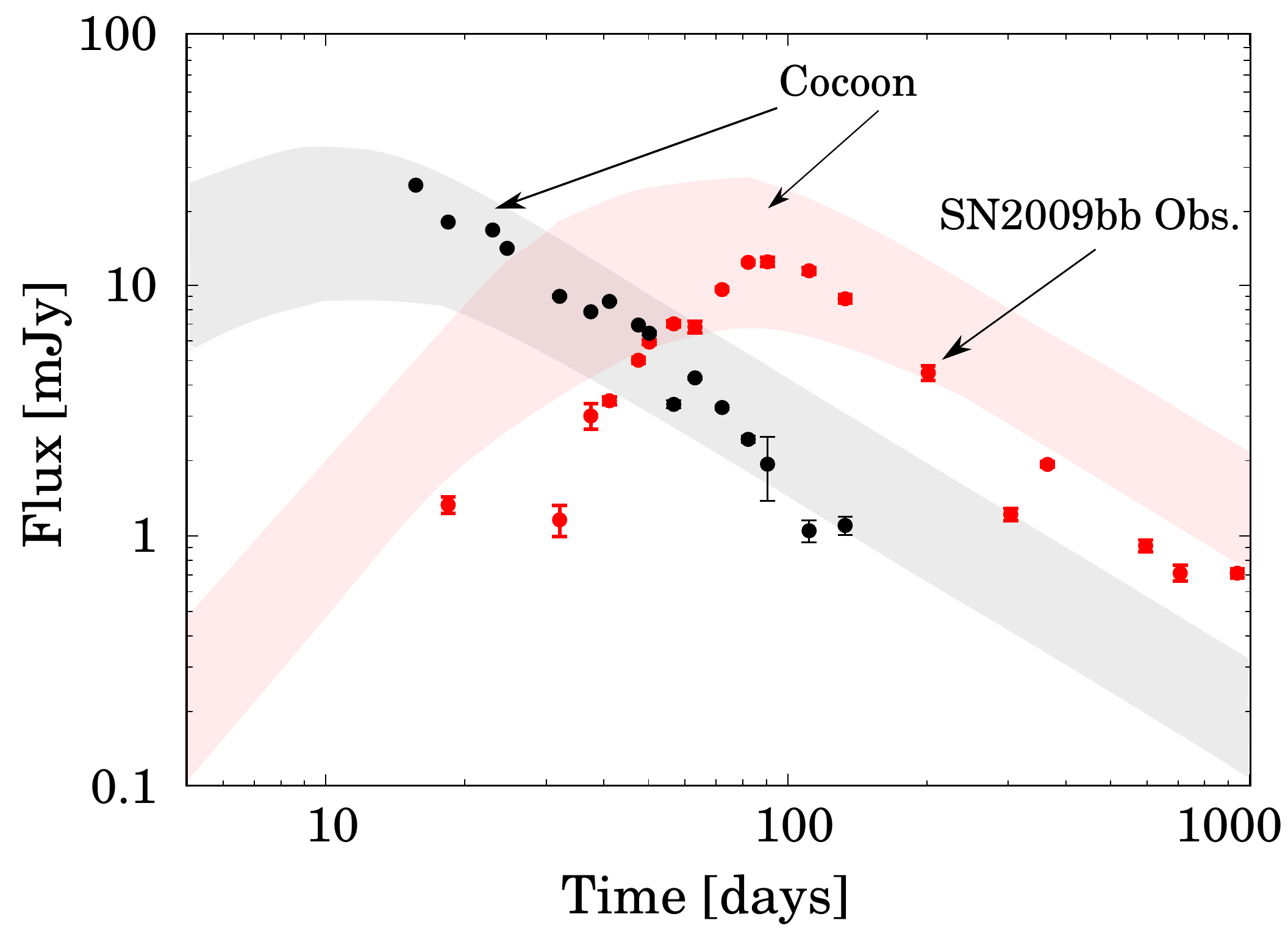}
\caption{Comparison between observed radio light curves of the SN 2009bb at 8.46 GHz (black dots, from \citealt{soderberg10}), 1.28/1.43 GHz (red dots, from \citealt{soderberg10, ray14}) and the radio emission from the cocoon computed by post-processing the results of the numerical simulations of a failed jet (pink and gray shaded areas, corresponding to 1.28 GHz and 8.46 GHz  respectively). The lower limit into the cocoon emission is computed for an observer located at $\theta_{\rm obs}=90^\circ$, while the upper limit for  $\theta_{\rm obs}=30^\circ$.
}
\label{fig5} 
\end{figure}

A comparison between the radio light curve of SN 2009bb and the cocoon emission produced by a failed jet (at $\theta_{\rm obs}=\pi/2$ the emission from the cocoon is similar in the two models) is shown in Figure \ref{fig5}. 
The observed radio emission shows a power-law decay as $\approx t^{-1.5}$, with large variability in flux possibly due to anisotropies in the progenitor wind.
The cocoon radio emission from a failed jet reproduces well the observed light curves, although it presents a slower decay with time, as $\approx t^{-1.15}$ (the other model where the relativistic jet breaks through the surface of the progenitor star successfully produces similar results).
A better agreement with the observation would be achieved if the average velocity (see Figure \ref{fig2}) had a faster decay in time, for instance by adjusting the density stratification of the circum-burst medium (which we have taken $\propto r^{-2}$) or by considering a different stellar structure.

We also computed the emission of a ``top-hat'' GRB jet observed at $\theta_{\rm obs} = \pi/2$ by using the result of simulations presented in \citet{decolle12b}. A GRB with isotropic energy $E_{\rm iso}=10^{53}$~erg  and $\epsilon_B = \epsilon_e = 0.1$ is is ruled-out from the observations, as it would peak at about 1000 days with a flux much larger than the one observed in the SN 2009bb. On the other hand, GRBs with a larger isotropic energy (see the discussion in Section \ref{secgrb}) or lower values of $\epsilon_B$ are not completely ruled-out by the data.

\begin{figure}
\centering
  \includegraphics[width=0.48\textwidth]{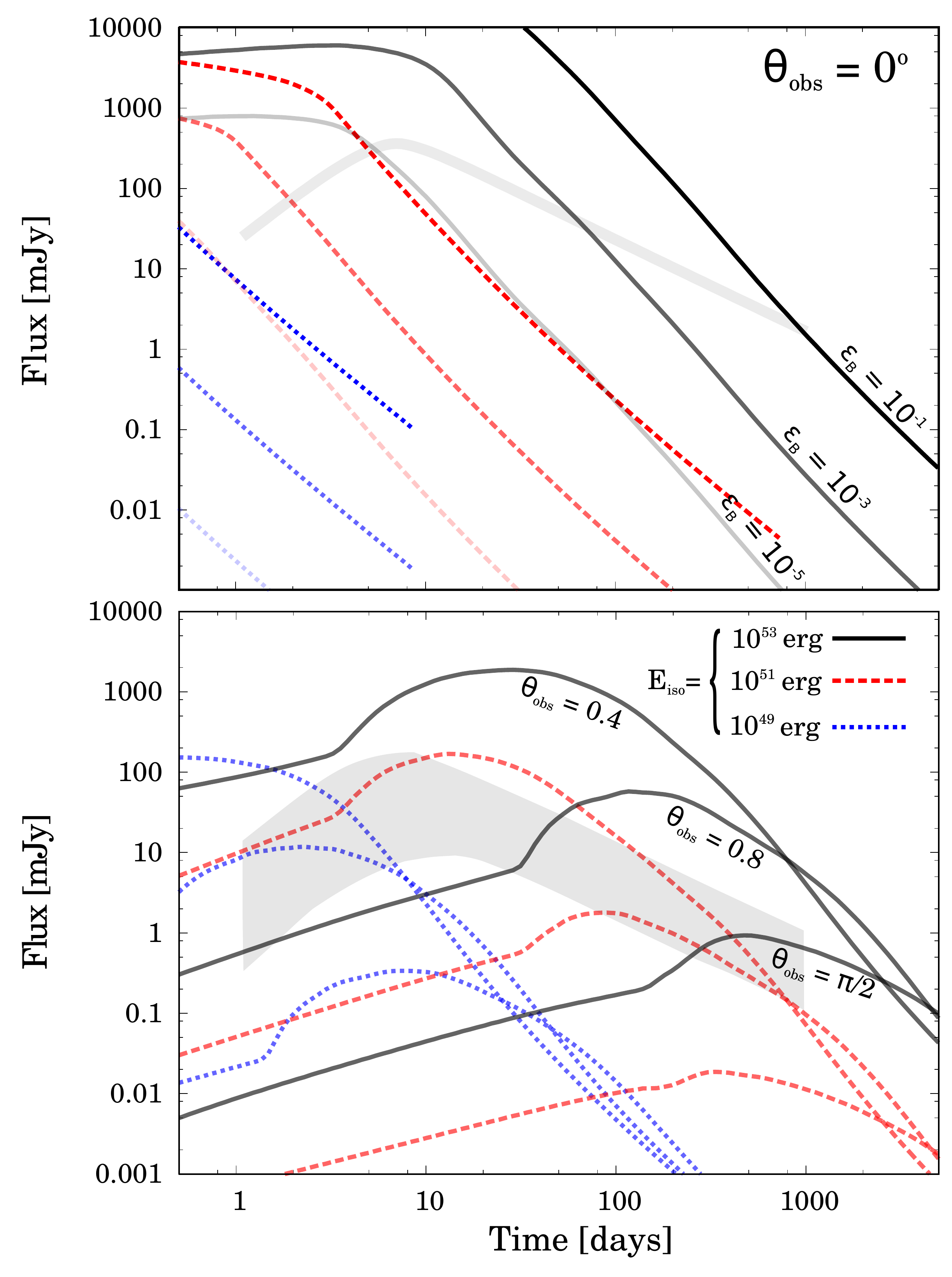}
\caption{
\emph{Top panel:} Comparison between the cocoon seen at on-axis (gray line) and the GRB radio afterglow emission (both computed at 8.46 GHz). The GRB afterglow emission is has been computed by using the simulations presented in \citet{decolle12b}, for a medium stratified as $\rho\propto r^{-2}$, with $\dot{M}=2\times10^{-6} M_\odot$ yr$^{-1}$, $\epsilon_e=0.1$, and $p=2.5$, for an observer located on- the jet axis (i.e $\theta_{\rm obs}=0^\circ$). The GRB afterglow has been computed for three isotropic energies (E$_{\rm iso}=10^{51}/10^{52}/^{53}$ erg (dotted blue, dashed red and full black lines respectively) and different values of $\epsilon_B$.
The cocoon emission dominates at large times if the energy and/or the value of $\epsilon_B$ is small. At small angles the jet always dominates the emission.  
\emph{Bottom panel}: Comparison between the cocoon (shadowed, gray area) and the GRB emission computed for off-axis observers located at $\theta_{\rm obs}=0.4,0.8,\pi/2$).
In this panel, $\epsilon_B=10^{-3}$.
If the GRB energy is low it is undetectable as the cocoon emission dominates. If the GRB energy is large, the cocoon can be detected at short times.
}
\label{fig6} 
\end{figure}

\subsection{GRB off-axis emission}
\label{secgrb}

While the first off-axis short GRB has been possibly recently observed associated to the GW/GRB170817A\footnote
{GRB170817 might have been an off-axis short GRB, or it might be the case that gamma-rays were instead produced by the shock break out of the cocoon through the neutron star  merger debris \citep[e.g.,][]{granot17, lazzati17, nakar18}}, observations of off-axis long gamma-ray bursts are still lacking. 
Type Ib/c SNe have been monitored for long time and they do not present evidence of a steeply rising lightcurve as one expects for an off-axis jet \citep[see, e.g.,][]{soderberg06, bietenholz14, ghirlanda14}.

Previous estimations of the off-axis emission from GRBs usually considered only the emission from the collimated jet. \citet{ramirez-ruiz02, nakar17, katir16}, among others, estimated analytically the accompanying emission from the cocoon and/or the SN. Figure \ref{fig6} shows under which conditions the cocoon emission can dominate with respect of the off-axis GRB emission. 
The figure shows that, when seen on-axis, the GRB emission always dominates at small times. The cocoon emission is important at late times unless the GRB isotropic energy and post-shock magnetic fields are large ($E_{\rm iso}\gtrsim 10^{53}$ erg,  $\epsilon_B \gtrsim 0.1$).
When observed on-axis in ``standard'' GRBs, the cocoon produces a flattening in the light curve. This has been studied in detail in the context of ``structured GRBs'' which are naturally generated by the presence of the GRB cocoon.

When observed off-axis (Figure \ref{fig6}, bottom panel), the cocoon dominates the 
radio emission at nearly all times if the GRB jet isotropic energy is small (i.e., $E_{\rm iso}\sim 10^{49}$ erg). Thus, a failed GRB/cocoon and the cocoon of a weak GRB observed off-axis will possibly produce similar emissions in radio. For an off-axis jet with large energy, the cocoon emission will dominate at small times, as the off-axis emission will peak at much larger times. 
For instance, the cocoon emission produces a peak at $\sim$ 10 days while a GRB with an isotropic energy of $10^{53}$ erg peaks at 100 days when observed at $\theta_{\rm obs}= 0.8$~rad.
In the case $\theta_{\rm obs} = \pi/2$, the observed time is $t_{\rm obs}= t - R \cos(\theta_{\rm obs})/c = t$. Thus, the jet radiation will contribute to the observed afterglow only when the jet has slowed down at nearly non-relativistic speeds and it becomes nearly spherical at $t_{\rm nr}\approx E_{\rm jet} v_w/\dot{M} c^3$. Thus, the peak will happen at a time $t_{\rm obs}\approx 1.76 \; (E_{\rm jet}/10^{51}{\rm erg})$~yrs.

We note that the presence of two peaks in Figure \ref{fig6} is due to the approximate treatment of the GRB jet considered here (i.e., a top-hat jet component treated separately with respect to the cocoon component). The computed GRB emission depends on the lateral expansion of the GRB jet, which will be strongly affected by the presence of the cocoon.
The lateral expansion of the jet would be slower in the presence of a cocoon that encapsulates the jet and provides pressure confinement.
Thus, the light curve will present a much smoother transition between the cocoon and the GRB peaks.


\section{Discussion and Conclusion}

In this paper, we have computed the non-thermal radio emission produced by the cocoon associated with a jet that propagates through the progenitor star but does not necessarily have enough energy to punch through the surface of the star.
About the X-ray emission, we expect that the radio cocoon emission will be similar for an observer located at 90 degrees. On-axis, the X-ray emission will be much larger for a successful jet due to the presence of highly-relativistic material. 

We do not try to perform a detailed fit of the observational data as the result would depend on the particular choice of our initial conditions. The radio emission, in particular, depends on the shock velocity (as a function of time and polar angle), the microphysical parameters, the energy in the cocoon and the density stratification of the circum-stellar medium. The energy in the cocoon in turn depends on several factors such as the angular structure and the opening angle of the jet and its magnetization parameter, and also on the stellar structure. 

\citet{decolle17} studied the quasi-thermal emission from the cocoon by employing two stellar models. They found that cocoon produced by the passage of jets through more extended stellar envelopes are less energetic and dimmer (the same effect should be present in the radio emission).
Also, we expect that large asymmetries in the jet (which should be studied by three-dimensional simulations), if present, would affect the cocoon velocity and energy.

A detailed comparison between observations and model could help constraining the jet characteristics and is left to a future study.

Our results show that the energy and velocity of the expanding cocoon, as well as the radio light curves and spectra are consistent with observations of relativistic supernovae, a sub-class of type Ic supernovae with mildly relativistic ejecta. 

Thus, our results strongly suggest that relativistic SNe have jets (failed or successful) that are at some large angle with respect to the observer line of sight.
The lack of detection of a GRB component both in radio (and X-ray) can be then explained by assuming that the GRB has a low energy, it is failed or it has a very large isotropic energy.

We showed that the cocoon from either a failed or a successful GRB has similar electromagnetic signatures when observed very off-axis. If the GRB is too faint to be detectable at early times in X-rays or at late times in radio, it is not be possible to distinguish between these two cases.

We also compared in detail (see Figure \ref{fig6}) the cocoon and the GRB radio emission when they are seen on-axis and off-axis (relative to the direction of propagation of the GRB). Our results clearly illustrate that the cocoon emission is going to be important in off-axis GRBs, and should be taken into account when making predictions of radio emission for future orphan afterglow surveys.

This also implies that all SNe driven by relativistic jets should present mildly relativistic material moving and emitting in radio unless the jet is choked in the deep interior of the star. 
Unless the SN is ejected before the GRB (with the disk surviving the SN ejection), the shock front of the SN, moving a highly sub-relativistic speed, needs much more time than the GRB jet to cross the star (as the energy for unit solid angle of the jet is much larger than that of the SN), and the cocoon will also arrive at the surface of the star before the SN shock does. The SN shock front, indeed, will then propagate through the cocoon. The implications on the SN dynamics and emission of the SN itself will be considered in a future paper.


\acknowledgments
We acknowledge useful discussions with Nissim Fraija, Diego L\'opez-C\'amara, Enrique Moreno-M\'endez, Raffaella Margutti, Enrico Ramirez-Ruiz and George Smoot.
FDC acknowledges support from the UNAM-PAPIIT grant IN117917. 
The simulations were performed on the Miztli supercomputer at UNAM as part of the project LANCAD-UNAM-DGTIC-281.



\end{document}